\documentclass[12pt]{article}

\usepackage{macros}

\usepackage{xcolor}

\usepackage{amsopn}

\usepackage{lmodern}
\usepackage{amsmath}

\usepackage{graphicx}

\usepackage{amssymb,amsmath}

\usepackage{relsize}

\usepackage{upgreek}

\usepackage{subfigure}

\usepackage{url}

\usepackage{graphicx}

\usepackage{algorithmic}
\usepackage{algorithm2e}

\pdfoutput=1
\usepackage{amsmath, amssymb}
\usepackage{graphicx,psfrag,epsf}
\usepackage{enumerate}
\usepackage{natbib}
\usepackage{url} 

\def\R{\mathbb{R}}

\def\X{\mathbf{X}}

\def\one{\mathbf{1}}

\def\C{\mathcal{C}}
\def\V{\mathbf{V}}
\def\bmu{\boldsymbol{\mu}}

\begin{document}

\def\spacingset#1{\renewcommand{\baselinestretch}%
{#1}\small\normalsize} \spacingset{1}


  \title{\bf ClusBench: The Clustering Benchmark Data Resource You've All Been Waiting For (?)}
  \author{David Hofmeyr
  \hspace{.2cm}\\
    School of Mathematical Sciences, Lancaster University}
  \maketitle

\bigskip
\begin{abstract}
Although some very common test beds exist for assessing the performance of clustering methods, large scale benchmarking is typically limited to relatively simplistic simulation set-ups. Here we describe the production and curation of close to 3000 synthetic data sets, derived from more than 200 publicly available data sets; the majority of which arose from real-world applications. By fitting a flexible non-parametric distribution to each base data set we are able to retain much of the nuance in real-world data which is difficult to reproduce in standard simulations, while also producing data sets whose sizes are sometimes substantially greater than the data sets from which they are derived. The synthetic data sets, plus an accompanying {\tt R} package, are available for download from~\url{https://github.com/DavidHofmeyr/ClusBench}.

\end{abstract}

\noindent%

\spacingset{1.45} 
\section{Introduction}
\label{sec:intro}

Assessing the quality of a clustering solution is made difficult by the unsupervised nature of the clustering task. Specifically, when faced with a clustering task in practice, there is no ``ground truth'' against which the discovered clusters can be compared, and thus validated. When extensive domain knowledge is available to the practicioner (either personally or through consultation), the reasonableness of a solution may be qualitatively assessed by investigating, e.g. the specific cluster allocations or the differentiating characteristics of different clusters through their summary statistics. However, the development of clustering methodology is typically not conducted under the assumption that such knowledge will be available, and moreover qualitative assessments do not lend themselves well to comparative evaluations across different clustering methods as general purpose tools.

\textit{Internal validation metrics} attempt to quantify the quality of a clustering solution by how it aligns with our commonsense interpretations of how clusters should be defined, with an overarching theme among these (and indeed many clustering models and algorithms themselves) being that observations which are more similar to one another should typically be clustered together while dissimilar points should typically be in different clusters. Even if we are able to commit to a precise meaning of similarity vs dissimilarity, almost inevitably such a metric will not be able to adequately validate all manner of cluster ``types'' (spherical, elongated, curved, inter-twined, etc.) which might exist, either separately or jointly in the same data set(s). As such they will end up favouring certain clustering models' solutions over others, purely because the objective of the clustering method aligns better with the definition of the validation metric and not necessarily because of an intrinsic superiority.

Simulation studies, in which synthetic data are generated from (typically mixtures of parametric) probability distributions, provide a useful framework for assessing clustering performance. Sampling from a mixture model may be seen as a two-step procedure in which first a mixture component (``label''), $Y$, is sampled from a discrete distribution $\boldsymbol{\pi}$ with cardinality equal to the number of components, and subsequently an observation, $X$, is sampled from the associated component distribution. These labels may then be used as a ``ground truth'' partition against which any clustering of the sampled observations can be compared, and numerous \textit{external validation metrics} have been developed to quantitatively assess the alignment between a clustering solution and the ground truth. However, the usefulness of simulation studies like these is limited by the difficulty of simulating data which are ``realistic'', in that they capture the intricacies of the sorts of data potentially seen in practice.

As a popular alternative, it is very common to rely on data from real applications for which a ground truth partition \textit{is} available, i.e. data sets more commonly used for evaluating classification methods. There are obvious limitations with such an approach, for instance the fact that the ``class distributions'' may not align with how we interpret clusters; either individually being split among multiple clusters of points or being distributed so that all clusters of points comprise observations from multiple classes, or a combination of these. Nonetheless it is arguably reasonable to assert that one clustering method is more useful than another as a general purpose tool if it is typically capable of producing solutions which align better with ground truth partitions, when these are available.\\
\\
Despite this view being fairly universally adopted, and comparative studies in much of the literature relying on these classification data sets, there is a dearth of large scale studies in which a very large variety of data sets is included. Frequently only a handful of data sets is included in any single study, and very few such studies explore more than a few tens of data sets. There is, as a result, an inevitable risk of biased comparisons due to ``cherry picking'' (often inadvertently), or developing clustering methods incrementally through experimentation with a particular collection of data sets, leading to tuning strategies which appear universal but are actually ``overfitting'' to the particular collection of data sets being explored; and not generalising well outside these.\\
\\
In an effort to provide a large and diverse set of clustering benchmarks, this paper describes the production and curation of close to 3000 synthetic data sets which are available for download from~\url{https://github.com/DavidHofmeyr/ClusBench}. We also describe the functionality of the {\tt R} package {\tt ClusBench}, also available from~\url{https://github.com/DavidHofmeyr/ClusBench}, which offers functionality to download and load these data directly into the {\tt R} environment, as well as the ability to interface with the methods used to produce them.

These data have been derived from a collection of more than 200 publicly available data sets for which ground truth groupings are available, and the vast majority of which are associated with real applications. These \textit{base data sets} were obtained from the Penn Machine Learning Repository database~\citep{pmlbr} and the UCI Machine Learning Repository~\citep{UCI}, which we accessed primarily using the {\tt R} packages {\tt pmlbr}~\citep{pmlbr} and {\tt ucimlrepo}~\citep{ucimlrepo}. To maintain the ``realism'' of the data distributions, while also allowing for the production of multiple varied data sets with similar characteristics and of size potentially vastly exceeding that of the some of the base data sets themselves, we fit a flexible probability distribution to each of them from which multiple samples could be drawn. From each of these distributions we then generated ten data sets, each of size 5000. In addition, for each base data set in which the class proportions are highly imbalanced, we also produced ten additional data sets of size 5000 where we modified the class priors to be equal for all classes. We then filtered out some batches of ten data sets by how clusterable they are, in the sense that clustering methods can reasonably be expected to produce a partition which reflects the classes substantially more than random guessing.

The reason for setting the size of each data set to 5000 is for the recognition that a large number of clustering methods require access to the full collection of pairwise distances, and this may produce a computational bottleneck for substantially larger data sets when evaluation is required over a large number of individual data sets. Moreover it is our experience that, unless the number of individual clusters is very large, 5000 is typically a sufficient number of observations to allow reliable ``detection'' of all clusters of reasonable magnitude.

For those wishing to experiment with data sets of different sizes, we make the following recommendations, in the interest of reproducibility:
\begin{enumerate}
    \item Users wishing to experiment with samples smaller than 5000 are advised to take the leading rows of each data set (up to the desired number of observations), so that results may be more comparable across different users' results.
    \item Users wishing to experiment with samples between 5000 and 50000 observations are advised to merge (subsets of) the total 10$\times$5000 = 50000 observations we have published from each base data set, and if possible document a deterministic merging method or make available any randomisation (e.g. indices in 1, ..., 50000 comprising each merged data set) so that others following the same experimental set-up may compare with and/or reproduce the results. 
    \item Users wishing to experiment with samples larger than 50000, or prefer to generate their own samples from the same or similar distributions, are advised to use the functionality in the {\tt ClusBench} package, which we describe in Section~\ref{sec:R}.
\end{enumerate}

\subsection{What The ClusBench Repository Is Not}

We have chosen to focus primarily on the ``realism'' of the data, while also standardising the testing context by sample size and number of samples from each base data set. As mentioned previously the sample size standard of 5000 points was chosen as we believe this is a threshold above which some clustering methods may be excluded due to computational demand, while the number of samples from each base distribution was set to 10 mainly to storage considerations. However, we recognise some important limitations, including:

\begin{enumerate}
    \item Big Data: The intention of this repository is not to support testing of the computational limits of the clustering task in any way. However, as alluded to previously, data sets of any size can readily be produced using the {\tt ClusBench} package.
    \item Task specific/robustness ``challenges'': There are existing clustering benchmark suites~\citep{FCPS, clustbench} which explore the important topic of ``stress testing'' clustering methods' robustness under archetypal contexts, including but not limited to (i) the presence of (and amount of) ``noise''; (ii) clearly quantified degree of class overlap; (iii) presence of (and number of) ``noise dimensions'' in which there is no class relevant information. We describe, in Section~\ref{sec:meta}, a partition of the data sets in {\tt ClusBench} which intersect with some of these aspects, but not in a well controlled manner as is achievable under simpler simulation set-ups.
\end{enumerate}

The rest of this paper is organised as follows. In Section~\ref{sec:sampling} we provide details on the procedure we used to convert a given data set into a probability distribution from which multiple samples of any size could be drawn. In Section~\ref{sec:filtering} we briefly discuss how we filtered out data sets deemed unreasonably difficult to cluster, in that producing a partition by any sensible clustering method is unlikely to align with the class distribution. With the final collection of data sets decided upon, we briefly explore their characteristics and describe a (hyper)clustering of the data sets in Section~\ref{sec:meta}. In Section~\ref{sec:R} we describe the main functionality of the {\tt ClusBench} package. We then give a brief concluding discussion in Section~\ref{sec:conclusions}

\section{From Data to Distribution}\label{sec:sampling}

The simplest way to obtain multiple data sets from a single benchmark is through resampling; effectively treating the data set as though it represents a population distribution \textit{a la} the non-parametric bootstrap. However, when the size of the base data set is relatively small it becomes challenging to produce much diversity across resamples. Moreover many clustering methods assume data are generated from a continuous probability distribution, which precludes the possibility of repeated observations.

The most natural solution is then to fit a continuous probability distribution to the data, and generate multiple samples from this. Depending on the data distribution this may be relatively straightforward; e.g. if the individual classes have distributions which are reasonably represented by parametric densities then the problem simply becomes one of parameter estimation. On the other extreme we may consider fitting a non-parametric density, either to the data as a whole or to each class separately, and sample from this. For the purpose of sampling, arguably the simplest non-parametric formulation is the Kernel Density Estimator (KDE), in which the sampling density is defined by the convolution of the empirical density (the simple bootstrap type resampling) with a smoothing \textit{kernel} (often a Gaussian density). Practically this is equivalent to first resampling directly from the data and then adding some additional noise, with the noise distribution defined by the smoothing kernel. This is highly principled, but choosing the right amount of noise is not always straightforward, and moreover even slightly too much noise can easily smooth over the subtleties in real-world data, and also obscure the separation of clusters. Between the two extremes lies the option of semi-parametric approaches, such as fitting a parametric mixture to each of the classes. However, although a Gaussian Mixture Model (GMM) may give a very good representation of each class, this instead introduces the potential that each class distribution possesses a large number of (possibly subtle) modes (sometimes many more than the actual number of modes) which an ``accurate'' clustering method may separate into different clusters.


In preparatory experimentation we encountered both the oversmoothing issue of KDEs and the ``extra-modes'' problem of GMMs very frequently, and no single existing strategy we explored was satisfactory. We therefore chose to use a modified KDE, 
in which first a ``shrunken'' version of the data is obtained by using a single iteration of a nearest neighbour Mean-Shift~\citep[MS]{meanshift} algorithm, before a weighted combination of the original data and its shrunken variant are resampled, before finally adding noise in the style of KDE sampling. Crucially the shrinking step accentuates the structure of distribution from a clustering perspective, by clarifying regions between high density clusters, so that the oversmoothing issue of KDEs is mitigated. Moreover, by defining the shrinkage in terms of the nearest neighbours of each observation we can also use this information to estimate the local covariance and so add smoothing noise which reflects this local variation, instead of the more common global \textit{bandwidth} applied in KDE. As a further modification we also first project the points from each class onto their leading principal components (PCs), so that the non-parametric resampling is conducted in a lower dimensional subspace. This is useful since the oversmoothing issue of non-parametric estimators is aplified in high dimensional settings, and the modified KDE we employ is not immune to this. For simplicity we model each class distribution in the orthogonal complement of its leading PCs with a Gaussian density.\\
\\
The complete set of steps taken to go from a data set to a sampling distribution is outlined below. It is worth noting that some of the decisions made (e.g. the proportion of variation to retain using PCA; the choice of ``imputation'' strategy; the number of neighbours; etc.) may be seen as somewhat arbitrary, however we point out that the intention is not necessarily to obtain the most accurate reflection of the ``true'' distribution underlying the data, but rather to obtain a probability distribution with similar characteristics to the data distribution, from which samples can be generated which fairly reliably capture the general structure of the distribution, in terms of the relative locations and shapes of the classes, and also \textit{some of} the subtleties and intricacies which differentiate real-world data sets from simpler simulations:
\begin{enumerate}
    \item Convert categorical variables: We used the standard approach of converting a categorical variable with $w$ categories into $w-1$ binary dummy variables. Note that the data loaded through the {\tt pmlbr} package do not directly indicate which variables are categorical, and so in order to automate the process we simply converted those variables with the fewest distinct values as though they are categorical.  
    \item Eliminate missingness: To handle missing entries we applied the following steps, in order:
    \begin{enumerate}
        \item We removed observations which had a missing ground truth label.
        \item We used an unsupervised Random Forest based imputation model to handle missingness in other variables, using the {\tt R} package {\tt randomForestSRC}~\citep{rfsrc}. \item If any missingness was not resolved using this model, we then removed variables which still contained more than 75\% missingness.
        \item Finally, if any missingness still remained we removed any incomplete observations.
    \end{enumerate}
    \item Define class-wise distributions: For each $c \in [C]$; where $C$ is the number of classes, let $\X_c \in \R^{n_c \times d}$ be the matrix of observations from class $c$. We then performed the following steps to arrive at a sampling distribution for the class:
    \begin{enumerate}
        \item Standardisation: If we let $\bmu^{(c)}$ be the mean of the rows of $\X_c$, and $\sigma^{(c)}$ the diagonal matrix with variable-wise standard deviations on the diagonal, then we set $\X'_c = (\X_c - \one\bmu^{(c)\top})(\sigma^{(c)})^{-1}$.
        \item PCA reduction: Let $\X^{'\top}_c\X'_c = \V\Lambda\V^\top$ be the standard spectral decomposition, and let $d_\tau = \min\{j| \sum_{i=1}^j \Lambda_{ii} \geq \tau \sum_{i=1}^d \Lambda_{ii}\}$, i.e. $d_\tau$ is the smallest number of dimensions needed to retain at least $\tau \times 100\%$ of the total data variation. We then set $d_c = \max\{2, \min\{50, d_{0.9}\}\}$, and project $\X'_c$ onto the first $d_c$ principal components, i.e. $\tilde\X_c := \X'_c\V_{1:d_c}$, where the subscript $1:d_c$ indicates the first $d_c$ columns of the matrix.
        \item ``Shrink'' $\tilde \X_c$ with nearest neighbours: For each $i \in [n_c]$ let $\mathcal{N}_{k_c}(i)$ be the indices of the $k_c$ nearest rows of $\tilde \X_c$ to row $i$ (we set $k_c = 2\lfloor \log(n_c)\rfloor$). Then let $\mathbf{A} \in \R^{n_c \times n_c}$ be the matrix with zeroes except in entries $\{(i, j)| j \in \mathcal{N}_{k_c}(i)\}$, where it takes the value one. Then the shrunken version of $\tilde \X_c$ is set equal to $\mathbf{Z}_c := \frac{1}{k_c} \mathbf{A}\tilde \X_c$.
        \item Define the sampling distribution for class $c$: To sample from the distribution characterising class $c$, first a point is sampled from the modified KDE. This involves three steps: (i) sample an index, $I$, uniformly from $[n_c]$; (ii) sample a location uniformly on the line segment joining the $I$-th row of $\tilde \X_c$ and the $I$-th row of $\mathbf{Z}_c$; and (iii) add location specific noise from a $N(\mathbf{0}, \Delta_I + \frac{1}{4}\Sigma_0)$ distribution, where 
        $\Delta_i; i \in [n_c]$ is the diagonal sample covariance matrix from the $k_c$ nearest neighbours of the $i$-th row of $\tilde \X_c$ and $\Sigma_0 = \frac{1}{n_c}\left(\tilde \X_c-\mathbf{Z}_c\right)^\top \left(\tilde \X_c-\mathbf{Z}_c\right)$ is a (full) shared covariance matrix related to the average amount of shrinkage from $\tilde \X_c$ to $\mathbf{Z}_c$. We can write this sample from the modified KDE as
        \begin{align*}
            X = \left(U \tilde \X_c + (1-U) \mathbf{Z}_c\right)^\top_{I:} + W_I,
        \end{align*}
        where $I$ is again the random index in $[n_c]$; $U \sim U(0, 1)$ and $W_I \sim N(\mathbf{0}, \Delta_I + \frac{1}{4}\Sigma_0)$, and we have used the subscript ``$I:$'' to mean the $I$-th row of the matrix.

        Now, $X$ lies in $\R^{d_c}$ as we only fit the modified KDE in the leading principal component subspace. We therefore then append $X$ with a $V \sim N(\mathbf{0}, \frac{1}{n_c-1}\Lambda_{(d_c+1):d})$, where $\Lambda_{(d_c+1):d}$ is the diagonal matrix of the trailing $d-d_c$ eigenvalues of $\X_c^{'\top} \X'_c$. After realigning $[X, V]$ with the cardinal basis, $V$ therefore accounts for the variation orthogonal to the leading PCs and retains the covariance of $\tilde \X_c$ in that subspace.
        After this realignment, the final step is to appropriately shift and re-scale to the appropriate scale and location of the observations in $\X_c$. The sample from the distribution for class $c$ is thus equal to $\sigma^{(c)}(\V [X, \ V] + \bmu^{(c)})$.
        
    \end{enumerate}
\end{enumerate}

\begin{figure}[t]
    \centering
    \subfigure[Satellite]{\includegraphics[width=0.95\linewidth]{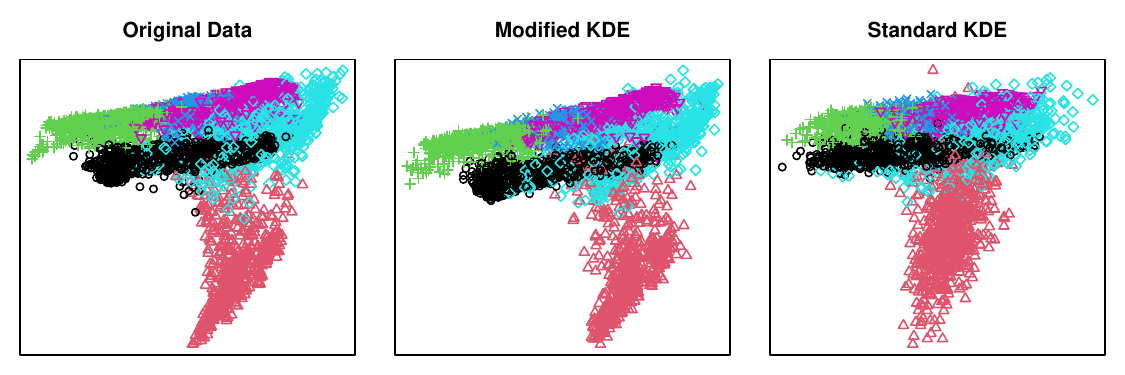}\label{fig:kdes1}}
    \subfigure[Auto]{\includegraphics[width=0.95\linewidth]{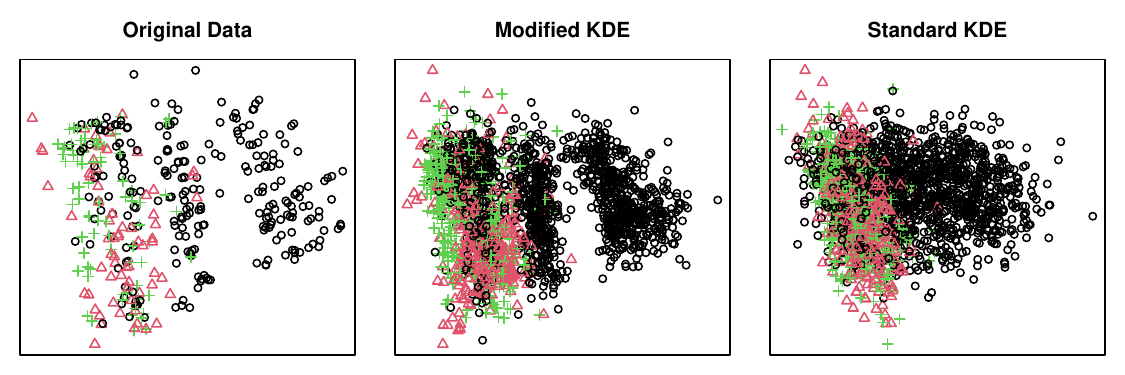}\label{fig:kdes2}}
    \caption{Two commonly used classification data sets (left), each with one synthetic sample generated using our modified KDE (middle) and one using a standard~KDE~(right)}
    \label{fig:kdes}
\end{figure}

We show two examples in Figure~\ref{fig:kdes}. In each case we show the first two principal component projections of (i) one of the base data sets to which we applied the above methodology; (ii) a sample generated using the proposed approach; and (iii) a sample generated by fitting a standard Gaussian KDE to each class using Silverman's rule of thumb bandwidth~\citep{silverman} of $\left(\frac{4}{n_c(2+d)}\right)^{\frac{2}{4+d}}\Sigma_c$. Figure~\ref{fig:kdes1} shows the thirty six dimensional Statlog Satellite data set~\citep{satellite} along with the two synthesised variants. The most obvious difference here is the substantial rounding effect of the standard KDE, rendering a sample whose classes appear close to Gaussian. The modified approach retains much more of the non-Gaussianity in the individual class distributions, although there is still some rounding of the especially elongated part of the class shown with \textcolor{red}{$\triangle$}'s. Figure~\ref{fig:kdes2} shows the seven dimensional Auto data set~\citep{auto}. In this case the original data set is quite small, with only 392 observations, and so being able to synthesise numerous and varied larger data sets, which respect the nuances of the distribution, becomes important. The multimodality in the class shown with $\bigcirc$'s leads to pathological oversmoothing with a standard KDE, which obscures all of the nuance in this distribution. On the other hand the modified KDE produces a data set which is pleasing as a representation of what we might expect a larger sample from the same underlying distribution to look like.

It is worth pointing out that, of course, choosing a smaller bandwidth will allow the standard KDE to retain more of the detail in the underlying distribution. However, it is notoriously challenging to select an appropriate global bandwidth which is small enough to retain detail but large enough to ensure samples do not contain ``almost exact'' replicates.

\section{Filtering Distributions/Data Sets}\label{sec:filtering}

When initially exploring data sets we capped the data dimensions at $n \leq 100000$ and $d \leq 2000$ (after conversion of categorical variables), and also ignored data sets which repeatedly produced errors even after modifying the automated process described above. We also sought to eliminate redundancy across the source repositories, but it is possible some remains.\\
\\
Once the initial collection of data sets, and their modified KDE distributions was obtained, it contained slightly under 250 in total. However, as we described previously it is not always the case that classification data sets lend themselves readily to assessing clustering performance, since if the class distributions do not align at all with what we interpret as clusters then it is unreasonable to expect any clustering methods to reliably distinguish classes. We therefore sought to filter out distributions whose samples could not reasonably be clustered in a way which reflects the actual classes.

\subsection{Quantifying Clusterability}

To limit subjectivity in selecting a clustering method to produce solutions for comparison with the actual classes, we applied all eight variants of the standard agglomerative hierarchical clustering algorithm implemented in {\tt R}'s {\tt hclust} function. The reasons for this choice include (i) across all eight linkage methods typically at least one can capture the clusters in any given data set reasonably well; (ii) despite their quadratic computational complexity, when all clustering solutions over quite a broad number of clusters are needed, their hierarchical nature makes finding the full collection of solutions comparatively efficient for data sets up to at least a few thousands; (iii) the pairwise distance calculations can be recycled, meaning the additional cost of computing all eight sets of clustering solutions instead of just one is relatively small; and (iv) aside from the number of clusters there are no additional hyperparameters for any of these methods, therefore further limiting subjectivity. 

\begin{figure}[h]
    \centering
    \includegraphics[width=0.95\linewidth]{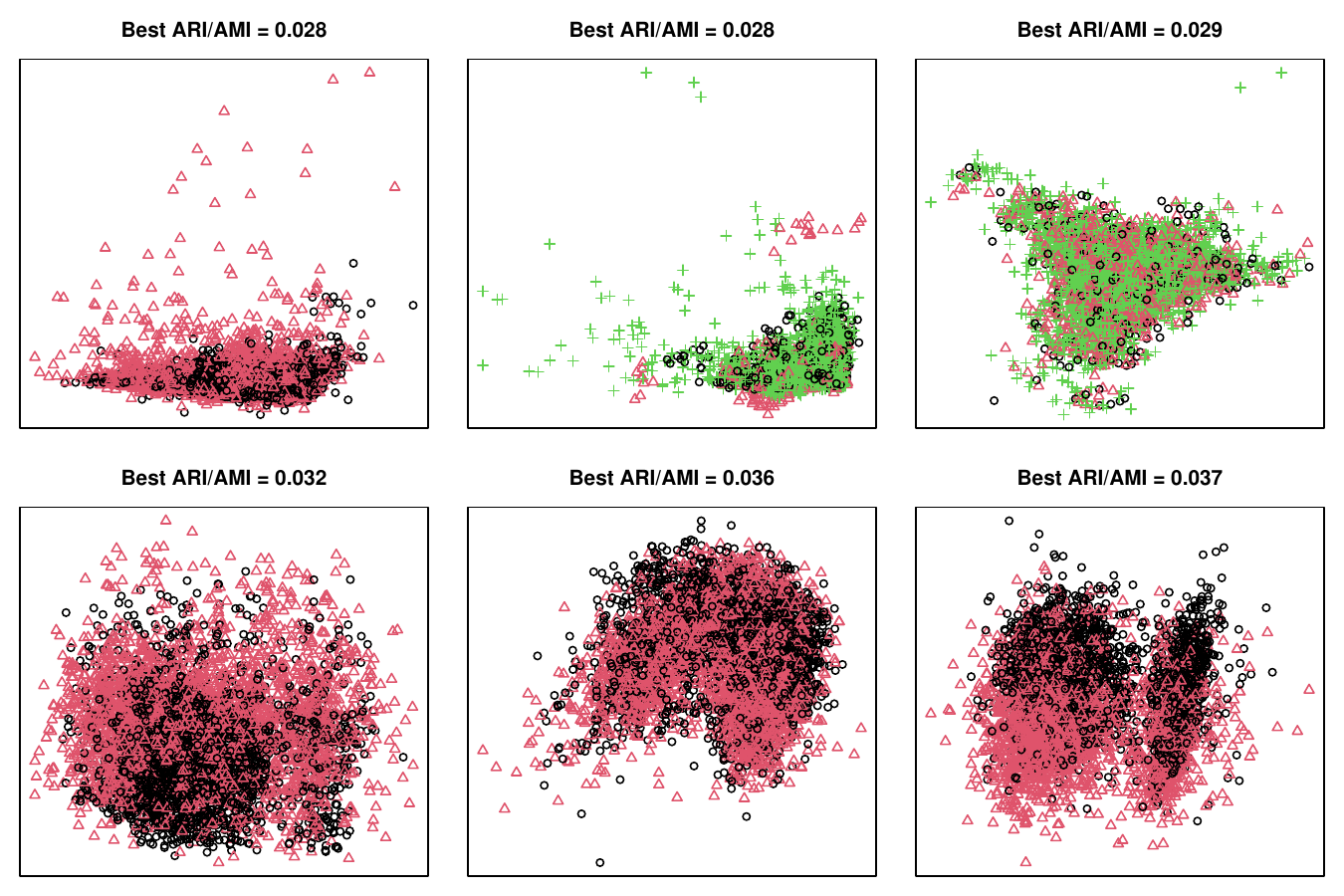}
    \caption{PCA projections of some of the data sets removed due to difficulty in obtaining clustering solutions which align with the actual classes. Points in different classes are differentiated by colour and point character. These may be seen as arguably the ``most clusterable'' of those removed.}
    \label{fig:unclusterable}
\end{figure}

We acknowledge that, although in what we argue accounts for the majority of typical scenarios at least one of the linkage methods is able to capture the clusters in a data set reasonably well, it is of course the case that this may not be true of \textit{all} data sets. We have therefore been conservative in removing data sets. In particular, for each data set we recorded, for each of the eight clustering models, the highest Adjusted Rand Index~\citep[ARI]{ARI} and Adjusted Mutual Information~\citep[AMI]{AMI} across a number of clusters from one up to three times the actual number of classes. The only exception to this is that for single linkage models we recorded the highest ARI and AMI up to ten times the number of classes, due to the model's sensitivity to noise meaning it sometimes ``devotes effort'' to merging singletons and small groups of points before merging substantial clusters. We then removed data sets only if none of the eight models achieved an average maximum ARI or average maximum AMI (when averaged over the ten data sets sampled from one of the fitted distributions) above $0.04$. As before this threshold is somewhat arbitrary, but visual inspection of data sets with low ``clusterability scores'', with simple PCA plots, suggested this as a reasonable threshold which does not lose instances which appear reasonably clusterable by eye. In addition the data which were removed showed, in almost all cases, at least one of the following characteristics: (i) no evidence of cluster structure at all; (ii) clusters which do not align with the class distribution; and (iii) extreme imbalance in class sizes. We show, in Figure~\ref{fig:unclusterable}, for each of the six base data sets with maximum ARI or AMI closest to 0.04 which we removed, one of the ten samples of size 5000 which were generated. Although the class proportions are not equal everywhere, and hence it \textit{could} be possible to find a spatial partition which gives a reasonable quality alignment with the classes, it is clear that when there are regions which are much more strongly associated with one class than others this typically does not align with how we interpret clusters.

\subsection{Rebalanced Class Proportions}

As mentioned above some of the data sets removed were challenging to cluster ``accurately'' purely because of highly imbalanced class proportions. This class imbalance phenomenon can be problematic even for the supervised classification task, but is particularly challenging in the clustering context. For each underlying data set in which the ratio of the largest to smallest class proportion exceeded three we therefore also generated ten data sets of size 5000 in which the sampling distribution was only modified by setting the class proportions all equal.

\begin{figure}[h]
    \centering
    {\includegraphics[width=0.66\linewidth]{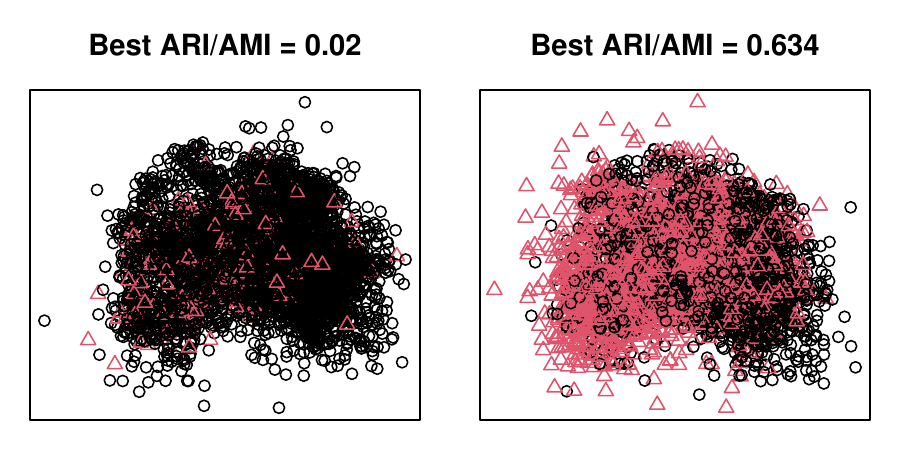}}
    {\includegraphics[width=0.66\linewidth]{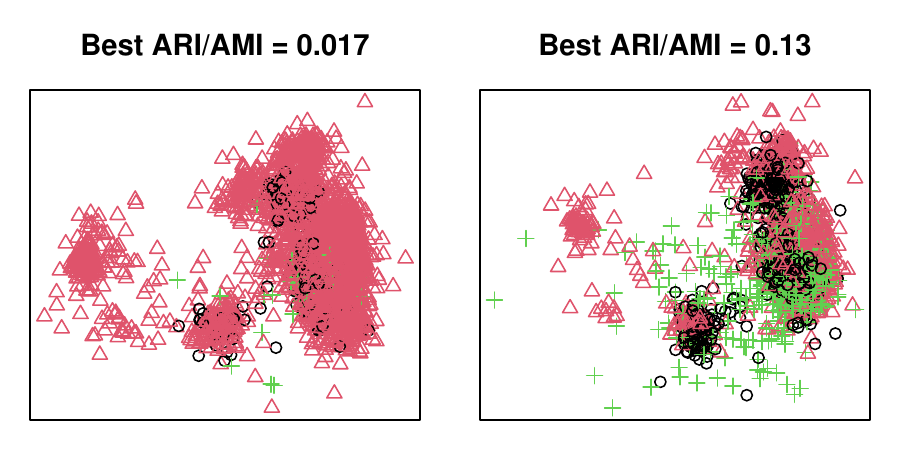}}
    \caption{Samples from two distributions which, after rebalancing of class proportions, become far more reasonably clusterable.}
    \label{fig:rebalanced}
\end{figure}

We show, in Figure~\ref{fig:rebalanced}, two examples where samples using the raw class proportions led to data sets which we removed due to ``unclusterability'' but which, after rebalancing the class proportions, crossed the 0.04 threshold for inclusion.

\section{Exploring the (Meta) Data}\label{sec:meta}

\begin{figure}[h]
    \centering
    \includegraphics[width=0.8\linewidth]{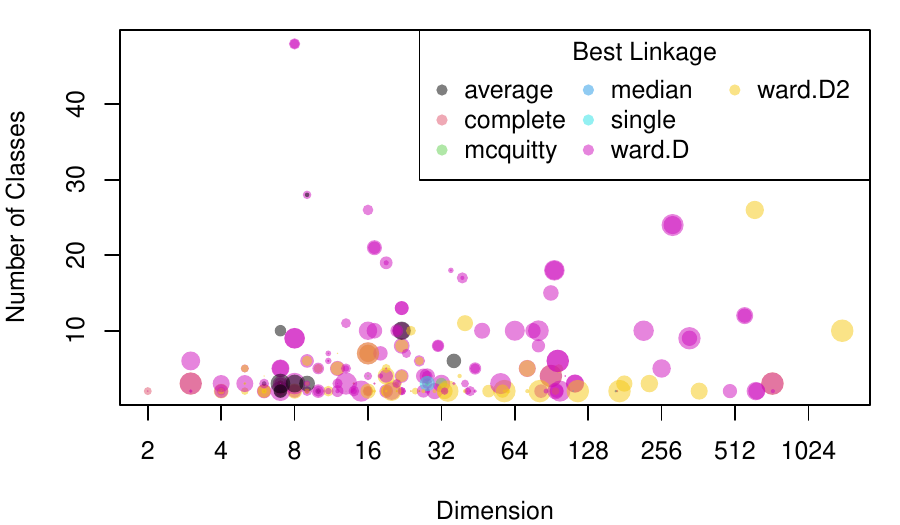}
    \caption{The 278 batches of ten data sets characterised by dimension and number of classes. In addition points are differentiated by size according to maximum clusterability, and by colour according to the hierarchical model giving the best result.}
    \label{fig:meta1}
\end{figure}

After filtering the batches of ten data sets generated from each of the sampling distributions (including using the raw proportions and the ``rebalanced'' proportions), a total of 278 remained. This final collection contains a large variety in terms of dimensionality; number of classes; class imbalance\footnote{note that although we used the ratio of largest to smallest class proportion in order to determine whether or not to also include a rebalanced variant of one of the fitted distributions, we quantify the imbalance going forward using the variance of the class proportions, as in~\cite{pmlbr}.}; and levels of clusterability. Figure~\ref{fig:meta1} shows the 278 (batches of) data sets in terms of numbers of dimensions and classes, with points differentiated by size according to the maximum clusterability\footnote{defined as the greatest ARI or AMI across all eight hierarchical models and across a range of numbers of clusters.} and by colour according to the linkage method in the hierarchical clustering which gave the best result. The two variants of Ward's linkage~\citep{ward} gave the best results far more often than any of the other linkage types, with average; complete and ``mcquitty'' each with a handful of best performances and almost no instances where one of single, median or centroid linkage is best. Within this collection very few of the data sets above about 50 dimensions have low clusterability (almost all small points in the plot lie towards the left). However, when considering any interaction with number of classes it is hard to discern anything with clarity partly because of the very large number of data sets with exactly two classes. It is worth pointing out, however, that the way we have quantified clusterability is in terms of the highest achievable ARI or AMI when assuming pseudo-oracle information regarding how many clusters gives the best result. It is arguably likely that appropriately selecting the ``best'' number of clusters is more challenging the more actual classes there are in the data, and so we may reasonably expect that the practical ``clusterability'' \textit{is} in fact related to the number of classes.

\subsection{Clustering the Data Sets}

\begin{figure}[h]
    \centering
    \includegraphics[width=0.65\linewidth]{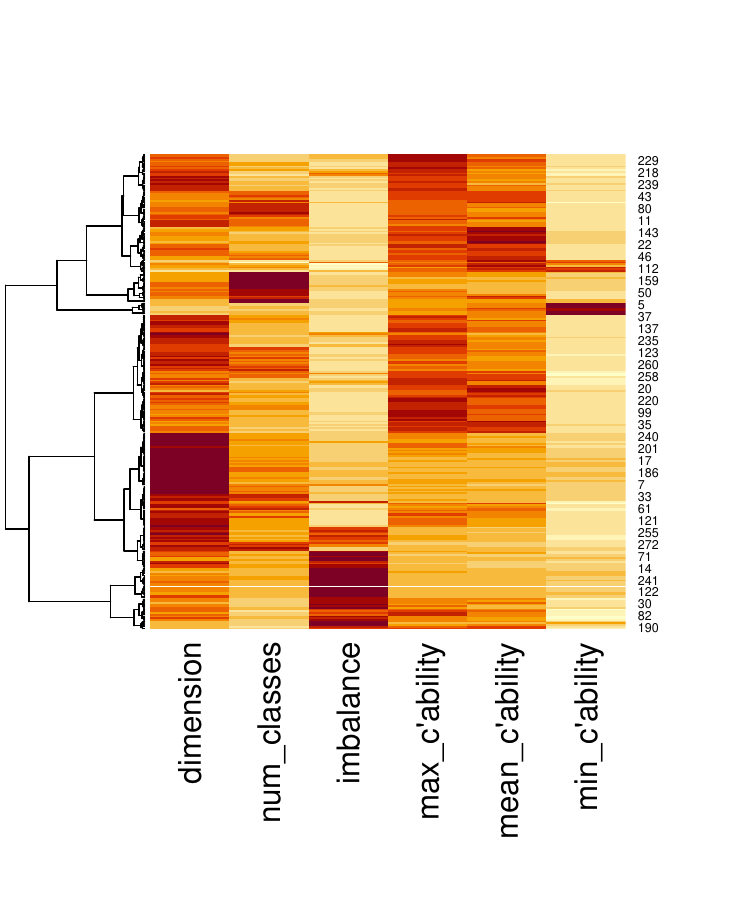}
    \caption{Batches of data sets differentiated according to dimension (on a log scale), number of classes, class imbalance and maximum; minimum; and mean clusterability (across linkage method). Cutting the dendrogram in the left margin at six clusters gives a reasonably balanced partition.}
    \label{fig:heatmap}
\end{figure}

Partly because of the large number of individual data sets in the repository, we sought to group them into data sets with different characteristics. In this way results obtained may be broken down by groups of data sets, and users may explore only certain groups based on the perceived strengths of the clustering method(s) being utilised. Figure~\ref{fig:heatmap} shows the data set characteristics in terms of dimension; number of classes; class imbalance; and maximum, mean, and minimum clusterability (by clustering model/linkage method). The dendrogram in the left margin is based on Ward type linkage but with Euclidean distance rather than squared Euclidean distance (i.e. {\tt method = "ward.D"} in {\tt R}'s {\tt hclust} function). Cutting this dendrogram to produce six clusters gives a reasonable balance of cluster sizes, with the exception of the resulting cluster 2, which is characterised by \textit{all} linkage methods producing fairly high quality clustering solutions (note the band of darkest colour in the rightmost column in the heatmap).

\begin{figure}[h]
    \centering
    \subfigure[Group 1: Many classes, medium clusterability]{\includegraphics[width=0.45\linewidth]{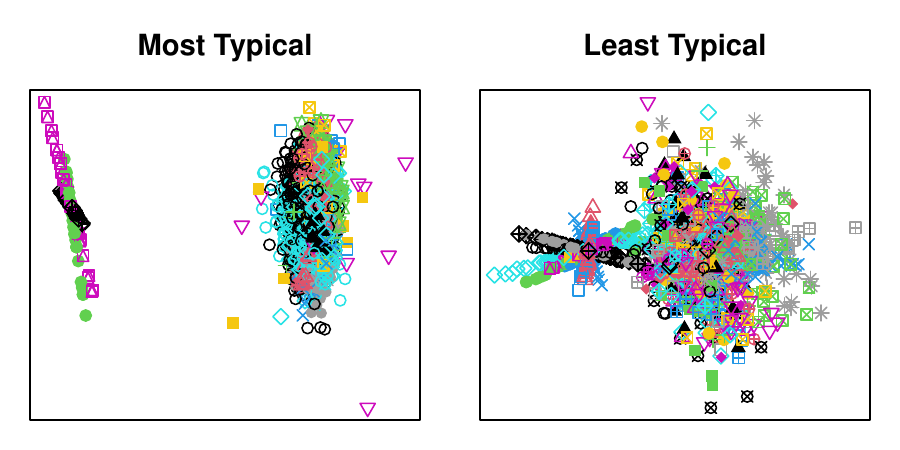}}
    \subfigure[Group 2: Low dimension, highest/easiest clusterability]{\includegraphics[width=0.45\linewidth]{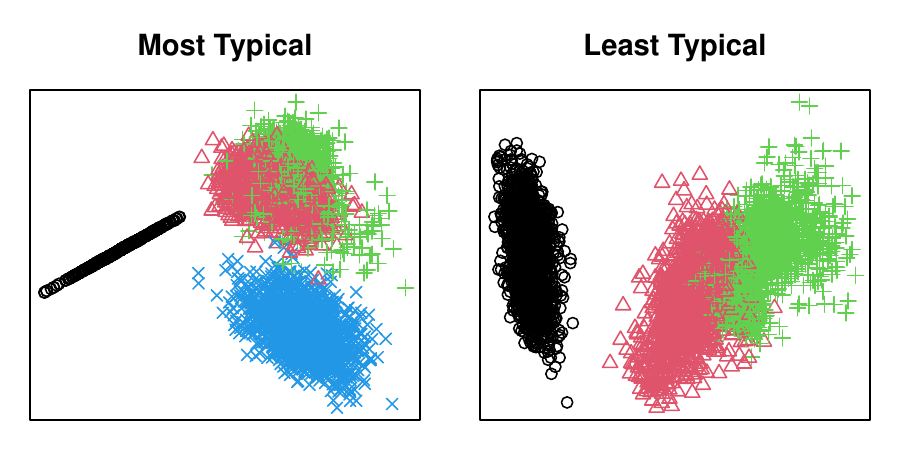}}
    \subfigure[Group 3: Lowest/most difficult clusterability]{\includegraphics[width=0.45\linewidth]{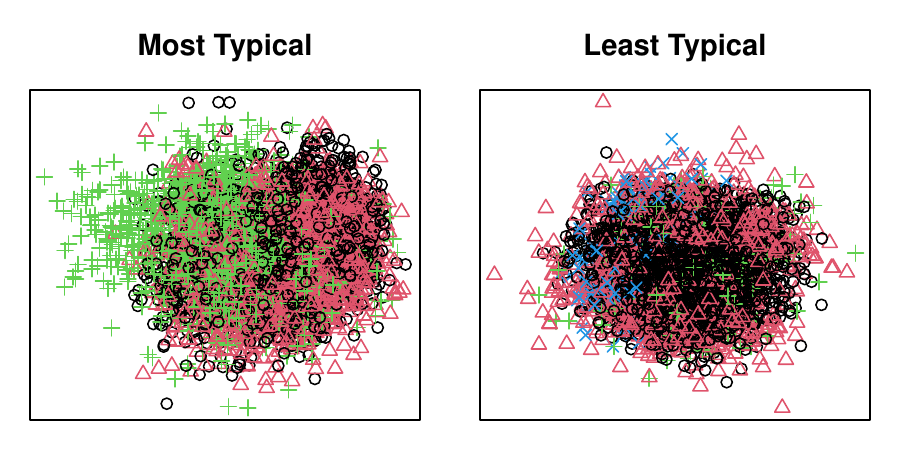}}
    \subfigure[Group 4: Few classes, medium clusterability]{\includegraphics[width=0.45\linewidth]{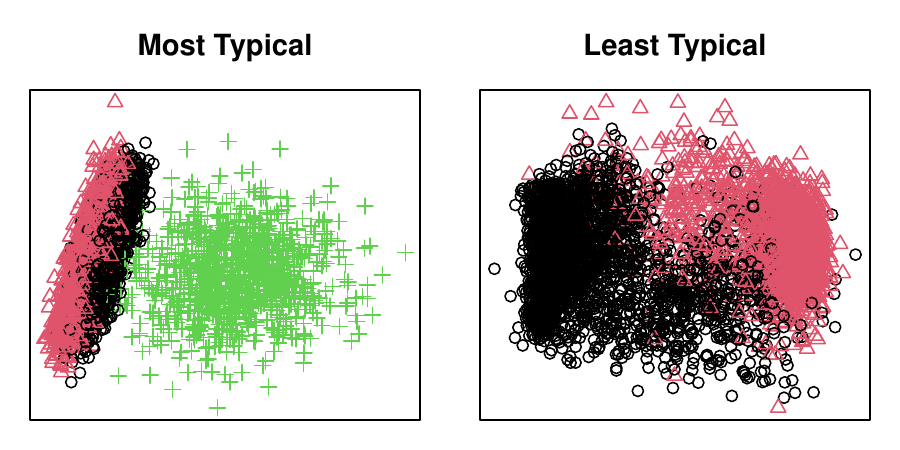}}
    \subfigure[Group 5: Highly unbalanced classes]{\includegraphics[width=0.45\linewidth]{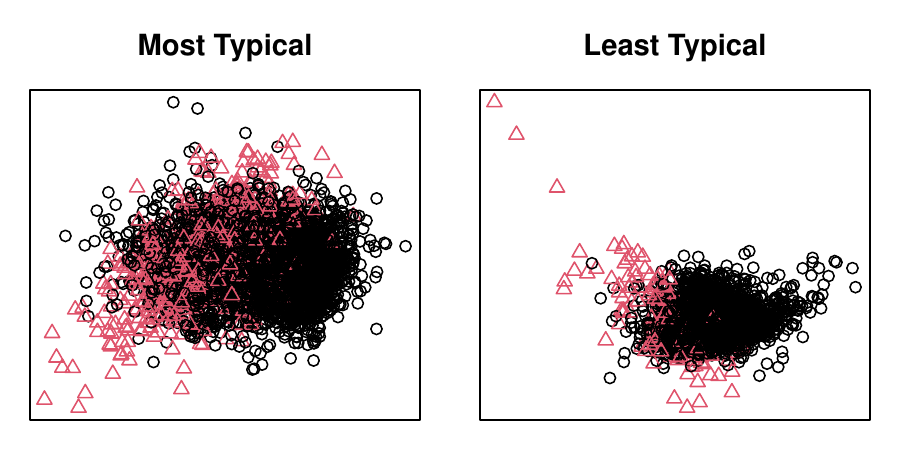}}
    \subfigure[Group 6: High/easy clusterability, mostly higher dimensional]{\includegraphics[width=0.45\linewidth]{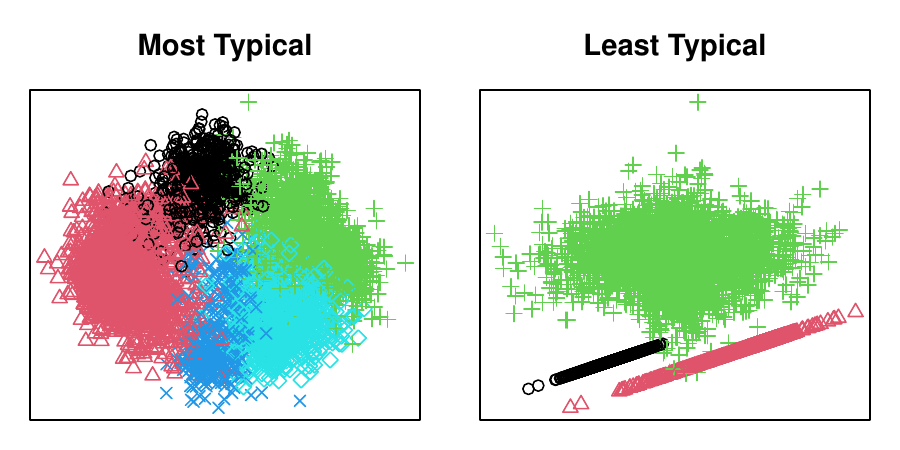}}
    
    \caption{Two data sets from each of the groups. In each case one of the samples from the ``most typical'' and ``least typical'' batch of ten is shown.}
    \label{fig:clusters}
\end{figure}

In addition, Figure~\ref{fig:clusters} shows two data sets from each group, projected onto their first two principal components. In each case the ``most typical'' member is taken to be that which has the smallest average distance from the rest of the group and the ``least typical'' is that with the largest average distance. In addition, a very brief description of some of the defining/discriminating characteristics of each group is given.

\section{The {\tt ClusBench} Package}\label{sec:R}

Here we very briefly describe the main functionality of the {\tt ClusBench} package. {\tt ClusBench} is a minimal package, only exporting five objects in total. These are:
\begin{enumerate}
    \item {\tt get\_data}: A function to download and import any of the data sets in the repository.
    \item {\tt sampleX}: A function to fit and sample from a modified KDE, as described in Section~\ref{sec:sampling}.
    \item {\tt SampleXy}: A function to fit and sample from a mixture of modified KDEs, stratified according to a given vector of class labels.
    \item {\tt dataset\_names}: A list of all 278 names for the batches of ten data sets in the repository. Each includes both the name of the base data set and a suffix {\tt \_original}, if the raw class proportions were used, or {\tt \_equal} if the rebalanced proportions were used for sampling. Note that not all data sets will have each of the suffices, since some class proportions were reasonably well balanced to begin with, while some highly unbalanced data were removed due to low clusterability.
    \item {\tt dataset\_statistics}: Additional information related to these 278 batches, including dimensionality; number of classes; class imbalance; clusterability scores and grouping as described in the previous section.
\end{enumerate}

By default {\tt get\_data} fetches the first of the ten samples of size 5000 for a given data set name, and others can be obtained by varying the argument {\tt version} in $1, 2, ..., 10$. For example, consider the very well known ``iris'' data set. The following code chunk loads the original iris data set from the package {\tt datasets} (part of {\tt R}'s base distribution), as well as two of samples of size 5000 from the {\tt ClusBench} repository. It then plots all three, projected onto their first two principal components.

\begin{verbatim}
R> data(iris, package = "datasets")
R> iris1 <- get_data("iris_original")
R> iris2 <- get_data("iris_original", version = 2)
R> par(mfrow = c(1, 3))
R> plot(princomp(iris[,1:4])$scores, col = iris$Species,
        main = "Original")
R> plot(princomp(iris1[,1:4])$scores, col = iris1$y,
        main = "ClusBench1")
R> plot(princomp(iris2[,1:4])$scores, col = iris2$y,
        main = "ClusBench2")
\end{verbatim}
\begin{figure}
    \centering
    \includegraphics[width=0.99\linewidth]{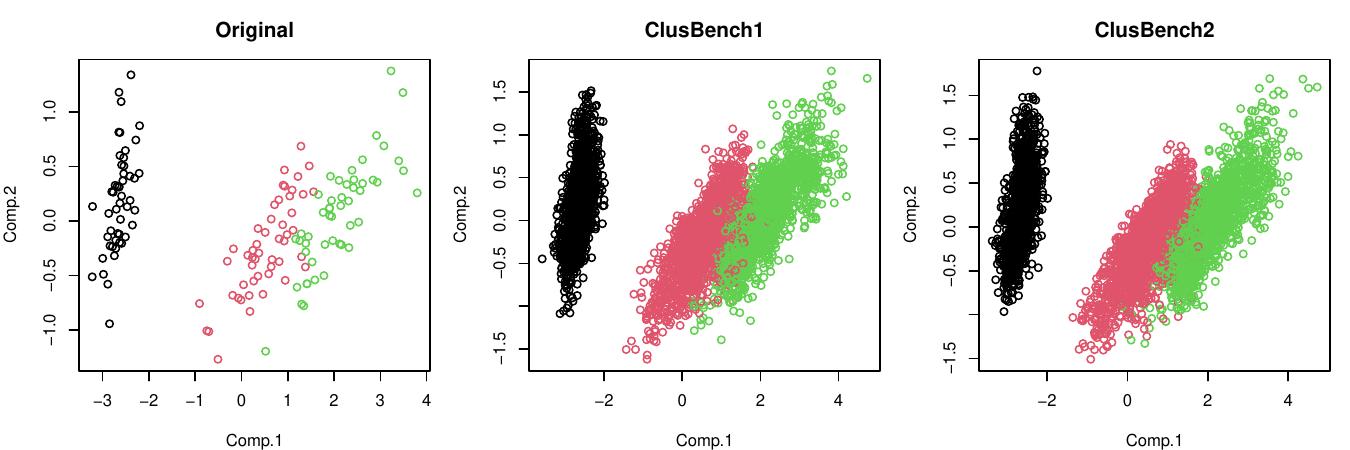}
    \caption{The iris data set, as well as two of the synthetic samples generated from it which are included in the ClusBench repository.}
    \label{fig:iris}
\end{figure}
The resulting plots are shown in Figure~\ref{fig:iris}. The individual classes in the iris data set are not far from Gaussian, and so this is not an especially interesting example except for the ubiquity of the iris data set making the generation of larger (and realistic) ``versions'' of it potentially interesting.

To illustrate the functionality of the functions {\tt sampleX} and {\tt sampleXy} we revist the Statlog Satellite data set. The following code chunk downloads and imports the original data set ({\tt satimage})\footnote{to run the following chunk you will need to first install the {\tt pmlbr} package.}, as well as one of the samples from the {\tt ClusBench} repository ({\tt satimage1}). In addition two new synthetic versions of the data set are created; one with {\tt sampleXy} and one with {\tt sampleX}. The first ({\tt sampleXy}) requires both the observations and the labels, and fits a modified KDE to each class separately) while {\tt sampleX} only requires the observations as it fits a single KDE to the entire data set. In order to retain the indices in the original sample, to be used to create labels for the resulting sample, the argument {\tt return\_ids} is set to {\tt TRUE}.

\begin{figure}[h]
    \centering
    \includegraphics[width=0.75\linewidth]{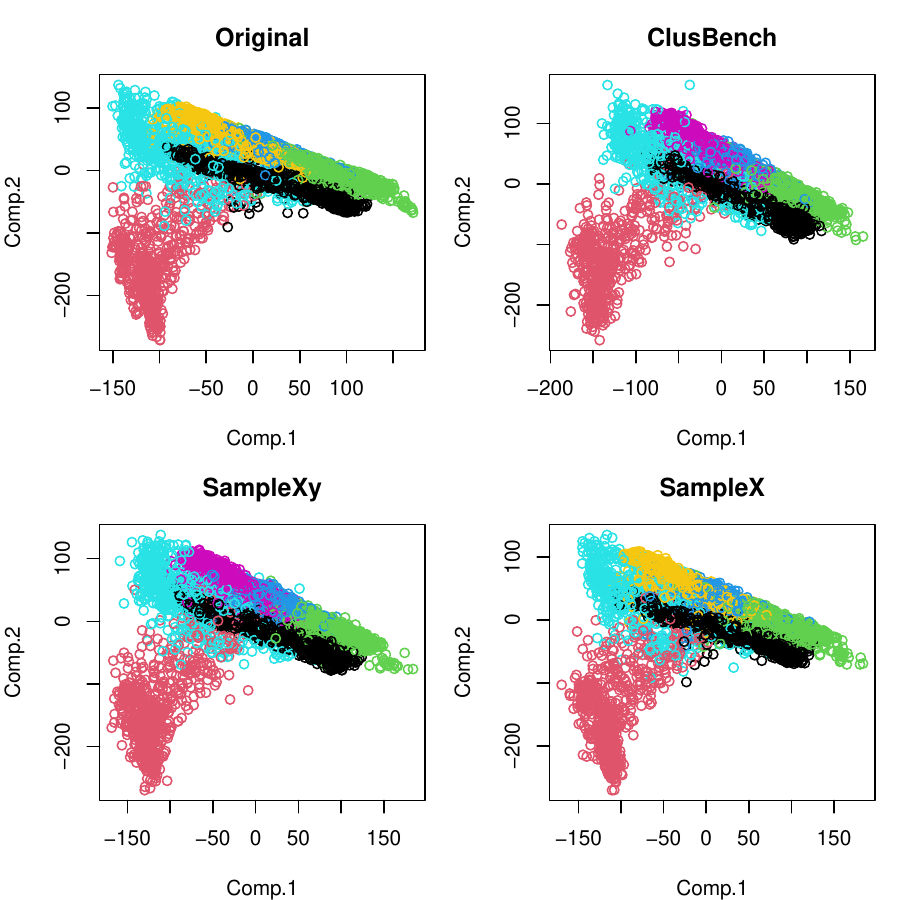}
    \caption{The ``satimage'' (Statlog Satellite) data set, as well as one of the synthetic variants in the ClusBench repository, plus two alternatives generated with the {\tt ClusBench} package.}
    \label{fig:satimage}
\end{figure}

\begin{verbatim}
R> satimage <- pmlbr::fetch_data("satimage")
R> satimage1 <- get_data("satimage_original")
R> satimage2 <- sampleXy(X = satimage[,1:36],
                    y = satimage$target, n = nrow(satimage))
R> satimage3 <- sampleX(X = satimage[,1:36],
                    n = nrow(satimage), return_ids = TRUE)
\end{verbatim}
Plotting the principal component projections of these four data sets, as in the following code chunk, gives the result shown in Figure~\ref{fig:satimage}\footnote{note that due to randomness in the sampling process, running the code will result in alternative versions of {\tt satimage2} and {\tt satimage3}.}.

\begin{verbatim}
R> par(mfrow = c(2, 2))
R> plot(princomp(satimage[,1:36])$scores, col = satimage$target,
        main = "Original")
R> plot(princomp(satimage1[,1:36])$scores, col = satimage1$y,
        main = "ClusBench")
R> plot(princomp(satimage2$X)$scores, col = satimage2$y,
        main = "SampleXy")
R> plot(princomp(satimage3$X)$scores, 
        col = satimage$target[satimage3$ids], main = "SampleX")

\end{verbatim}

\section{Conclusions}\label{sec:conclusions}

We described the production and curation of a large collection of benchmark data sets for assessing the performance of clustering methods. By fitting flexible non-parametric distributions to more than 200 \textit{base data sets}, we were able to generate multiple samples from each of an extremely varied collection of data generating mechanisms, which retain much of the nuance of real-world data which is very difficult to replicate in standard simulations. We filtered out some data sets which may be seen as representing unrealistic clustering tasks, in that we should not expect a clustering method to be able to partition the data in a way which reflects the actual classes in the data. We also described the functionality of an {\tt R} package which can be used to download and import the almost 3000 data sets directly into {\tt R}, as well as to interface directly with the methods used to produce them and so generate additional data sets for further exploration and testing of clustering capabilities.

\bibliographystyle{plainnat}

\bibliography{Bibliography-MM-MC}
\end{document}